\documentclass[pre,aps,preprint,showpacs,showkeys,nofootinbib,
endfloats]{revtex4-1}
\usepackage[dvipdfm]{graphicx}
\usepackage[dvipdfm]{rotating}
\usepackage{epsfig}
\usepackage{subfigure}
\usepackage{amsmath}
\usepackage{amsfonts}
\usepackage{amssymb}
\usepackage{color}
\usepackage{ulem}

\begin{document}
\title{Equivalence between  microcanonical  methods for lattice models}
\author{Carlos E. Fiore}
\affiliation{Departamento de F\'{\i}sica, Universidade Federal do Paran\'a \\
Caixa Postal 19044, 81531-000 Curitiba, Paran\'a, Brazil}
\author{Cl\'{a}udio J. DaSilva}
\affiliation{Departamento de Matem\'{a}tica e F\'{\i}sica, Pontif\'{\i}cia 
Universidade Cat\'{o}lica de Goi\'{a}s,\\ 74605-010, Goi\^{a}nia, Goi\'{a}s, 
Brazil}
\date{\today}

\begin{abstract}
The development of
reliable methods for estimating microcanonical averages constitutes
an important issue in statistical mechanics. 
One possibility consists of  calculating a given 
microcanonical quantity by means of typical relations in the grand-canonical 
ensemble.
But given that distinct ensembles are equivalent only at the
thermodynamic limit, a natural question is if finite size effects
would prevent such procedure.
In this work we investigate thoroughly this query in different systems yielding 
first
and second order phase transitions.
Our study is carried out from the direct
comparison  with the thermodynamic 
relation $(\frac{\partial s}{\partial e})$, where
the entropy is obtained from the density of states. 
A systematic analysis for finite sizes is undertaken. 
We find that, although results become inequivalent
for extreme low system sizes,   the equivalence holds true
for rather small $L$'s. Therefore direct, simple (when compared with other well
established approaches) and very precise microcanonical quantities can
be obtained from the proposed method. 
 

\end{abstract}

\maketitle

\section{Introduction}
The development of efficient Monte Carlo (MC) methods constitutes
 a key problem in statistical mechanics.  
Typically, numerical simulations are  performed by  
one-flip algorithms that generate a grand-canonical ensemble, in 
which  intensive quantities are held fixed \cite{landaubook}. 
However, in some cases,  a different ensemble
would be more appropriate. For example, strong first-order
 phase transitions become extremely hard
to simulate by one-flip grand-canonical schemes:
the presence of different phases separated by large free-energy
barriers makes the system to be trapped into metastable
states  at the phase coexistence, even for small
system sizes. Although different grand-canonical procedures
have been proposed \cite{bouabci,berg,fioreprl}, the microcanonical 
ensemble is also an appropriate way to circumvent  these problems. 
In this case, 
the simulation is carried out for fixed energies and 
 intensive thermodynamic quantities are treated as external variables, 
hence avoiding large entropic barriers.

Different microcanonical schemes have been proposed in the last years. 
Entropic sampling  \cite{lee}, broad histogram method \cite{pmc} and the  
Wang-Landau (WL) method \cite{wang}  are some examples of
procedures in which the  density of states (DOS) is estimated numerically. 
Other  microcanonical approaches, not requiring
the knowledge  of the DOS, have also been developed.
In  such cases, the temperature and other intensive quantities 
are obtained from  auxiliary relations.
For example, 
Creutz \cite{creutz} has generated a microcanonical ensemble by assuming 
a canonical distribution of the energies carried by a ``demon'', where
the total energy is not strictly conserved, but it fluctuates
above a constant lower bound.
More recently, Martin-Mayor \cite{mayor} proposed a method where
the temperature is obtained from an  ensemble of fixed energy 
(including the kinetic energy). The 
temperature is calculated from the fluctuations of the spin part.

To analyze the discontinuous
transitions in the Blume-Emery-Griffiths (BEG) \cite{BEGMODEL} and
asymmetric Ising models \cite{wang2}, 
it has been employed a ``trick'' \cite{fiore6}, where intensive quantities 
 are calculated through  expressions originally derived in the 
grand-canonical ensemble.
Although the equivalence  is granted  in the thermodynamic limit, 
 Gibbs ensembles may be inequivalent 
for finite systems, including  short range interactions
at the phase coexistence \cite{gross}. 
Therefore, in order to the protocol can be extended to more
general situations, one should test under what conditions 
the finite size effects  hinder the equivalence between methods.
In other words, it should be  verified that  
 intensive quantity   as a matter of fact correspond to the genuine
microcanonical temperature (obtained from the derivation 
of entropy with respect to the energy).  

The first goal of this paper is then to answer 
the above query. For so,   
different aspects of the method  will 
be exemplified by means of distinct 
 lattice systems  yielding  first and second-order phase
transitions. 
We first address the Ising model, for which the DOS are known
exactly \cite{beale}. 
Then we consider as next examples the Blume-Capel (BC) 
and the Potts models. Although they do not
present exact DOS, we are going to compare with the very 
 efficient WL sampling as a benchmark.
 The BC model is an interesting case, since  its DOS
has been obtained by performing a random walk in the space of two parameters 
(in similarity to several lattice models presenting distinct particle 
interactions) \cite{claudio,wang2}. 
We intend to verify if  the calculation of the temperature
will be changed by different restrictions in the random walk 
(as performed in Ref. \cite{claudio}). 

The Potts models is also  a very interesting test.  Unlike the
previous cases, its discontinuous transitions (yielding for  $q>4$) 
presents a genuine microcanonical feature, the existence of a loop 
\cite{lee,mayor,komura}.
Thus, it is important to verify if our approach
not only reproduces this remarkable signature but also is equivalent
to the $(\frac{\partial s}{\partial e})$.
In addition, we are also exploiting the interesting $q=4$ case, that 
although yielding a continuous transitions, it possesses distinct behaviors 
including  logarithmic scaling corrections \cite{mayor,fernandez,sokal} 
and  a double peak probability distribution \cite{fukugita}. 
As it will be shown further, the methods becomes equivalent  
in  all above models for 
relatively small system sizes $L$. The equivalence  includes
not only the  temperatures but also
all extrapolated 
thermodynamic limit  points. However, for extreme small
$L$'s, the results do not agree and thus the intensive quantity 
can not be recognized as the thermodynamic temperature. We will present a
 detailed analysis showing how the methods converge when $L$  increases.

A second contribution here is to exploit the advantages. 
Besides its low computational cost, it does not require
criterion for achieving convergence of results.
Other immediate advantage is that intensive 
quantities are evaluated directly from standard numerical simulations and  
 become more precise as  the system size increases. In
addition, the method is very easily extended for other lattice models.

This paper is organized as follows: In Sec. II we review 
the methods for calculating the temperature. In Sec. III 
we show the numerical results for the   models and in Sec. IV 
we present our conclusions.

\section{Microcanonical Temperature}

For  a given system size $L$ and  energy per site $e=E/V$ (where
$V=L^{d}$ and $d$ is the dimension), the  inverse of microcanonical temperature
$\beta_L$ is  obtained through the expression
\begin{equation}
\beta_L=(\frac{\partial s}{\partial e})_L,
\label{twl}
\end{equation}
where $s=s(e,L)=\frac{k_B}{L^{d}}\ln \Omega(E,L)$ is the entropy per site
 and $\beta_L=1/k_{B}T_L$. The quantity $\Omega(E,L)$ denotes the DOS for 
given $E$ and $L$. For the Ising model, the  $\Omega(E,L)$ is known exactly
\cite{beale}. For the  other models, we shall estimate $\Omega(E,L)$ using the
WL sampling \cite{wang}. 

The WL sampling is a  powerful technique to calculate $\Omega(E,L)$ by 
carrying out a random walk in energy space with an acceptance probability
proportional to $1/\Omega(E,L)$, i.e.,
\begin{equation}
 P(E_i\rightarrow E_j)=\min\left[ \frac{\Omega(E_i,L)}{\Omega(E_j,L)},1 \right],
\end{equation}
where $E_i$  and $E_j$ are the energies of the current and a possible new
configuration, respectively. For each new accepted configuration an energy
histogram $H(E)$ is accumulated.

During the random walk, whenever a move to a configuration with energy $E$ is
accepted, $\Omega(E,L)$ is updated by multiplying it by a ``modification
factor'' $f>1$ that accelerates the diffusion of the random walk, and an unit is
added to the histogram $H(E)$. The initial choice of $f$ is $f_0 = e =
2.71828\dots$. $\Omega(E,L)$ is multiplied by $f$ until the accumulated
histogram $H(E)$ becomes flat. We then reduce $f$ by setting $f\rightarrow
\sqrt{f}$, and resetting $H(E) = 0$ for all energy values. The simulation
converges to the true value of $\Omega(E,L)$ when $f$ approximates to the unit. 
In particular, in this work we use the improved Wang-Landau sampling
proposed by Cunha-Netto {\it et al.} \cite{cunha}. Their approach use adaptive
energy windows to eliminate border effects that affect the density of states
mainly of q-states Potts model, which is our case here. In our 
simulations the criterion of flatness was taken as each value of the 
histogram reaching at least $80$\% of the mean value $\langle H(E)\rangle$ 
for the BC model and $90$\% for the Potts model. The histograms are 
generally checked after each $10000$ MC steps. Here we performed $10$ different
runs for the same $L$ with different initial seeds in order to reduce
statistical fluctuations. For the BC model, the DOS is obtained by performing a
random walk for two parameters $E$ and $E_2$ ($E_2=\sum_{i}\sigma_{i}^{2}$).
An  immediate advantage of the WL is that a single run gives the
DOS for the whole range of energy, which provides the calculation of
canonical averages for any temperature.

Now, following Ref. \cite{fiore6} we proceed to obtain the  temperature
$T$ with respect to the microcanonical ensemble.  The method consists in writing
down the probabilities of different microscopic configurations in the
grand-canonical ensemble and further resorting the equivalence of ensembles.

The probability distribution $P(\sigma)$ of a microscopic configuration $\sigma=(\sigma_1,\sigma_2,...,\sigma_V)$ in the grand-canonical
ensemble is given by $P(\sigma)=\exp\{-\beta{\cal H}(\sigma)  \}/\Xi$, 
with Hamiltonian ${\cal H}(\sigma)$  reading
\begin{equation}
{\cal H(\sigma)}=-J\sum_{(i,j)}\sigma_{i}\sigma_{j},
\label{1}
\end{equation}
for the Ising model and
\begin{equation}
{\cal H(\sigma)}=-J\sum_{(i,j)}\sigma_{i}\sigma_{j} +\Delta \sum_{i}
\sigma_{i}^{2},
\label{2}
\end{equation}
for the BC model and
\begin{equation}
{\cal H(\sigma)}=-J\sum_{(i,j)}\delta_{\sigma_{i},\sigma_{j}},
\label{3}
\end{equation}
for the Potts model. Parameters $J$ and $\Delta$  are the energy 
between two nearest-neighbor spins and  the crystalline field,
respectively. The spin variable $\sigma$  
takes the values $-1$ or $+1$ for the Ising model, $-1,0$ or $+1$ 
for the Blume-Capel and $0,1,...,q-1$, for the Potts model. In 
all cases, the summations are restricted over nearest neighbor sites. 

By considering the transition $-1\leftrightarrow 1$ for
Ising and BC models and denoting $\sigma^{k}$ by a microscopic configuration 
which differs from $\sigma$ only by the value of the spin at the site $k$, that
is, $\sigma^{k}=(\sigma_1,\sigma_2,...,-\sigma_{k},..,\sigma_V)$, the 
ratio between  $P(\sigma)$ and  $P(\sigma^{k})$ in the grand-canonical 
ensemble is given by
\begin{equation}
\frac{P(\sigma)}{P(\sigma^{k})}
= \exp \{ 2 \beta \sigma_{k}[\phi_k(\sigma)]\},
\label{14}
\end{equation}
where
\begin{equation}
\phi_k(\sigma) = J\sum_{\delta} \sigma_{k+\delta},
\label{15}
\end{equation}
whose summation is performed over $\delta$ nearest neighbor sites. 
If the average of an arbitrary state function in the 
grand-canonical ensemble is given by $\langle \
f(\sigma)\rangle_{gc}=\sum_{\sigma}f(\sigma) P(\sigma)$, 
from  Eq. (\ref{14}) we have that
\begin{equation}
\langle f(\sigma) \rangle_{gc}=\langle f(\sigma^{k})
\exp \{ 2 \beta \sigma_{k}[\phi_k(\sigma) ]  \}\rangle_{gc}.
\label{e16}
\end{equation}
By taking Eq. (\ref{e16}) for $f(\sigma)$ given by  
$f(\sigma)=\delta(\sigma_k,+1) \delta(\phi_{k}(\sigma),{\bar E})$,
we have that 
\begin{equation}
  e^{-2\beta {\bar E}}= \frac{\langle\delta(\sigma_k,-1)
\delta(\phi_{k}(\sigma),{\bar E})  \rangle_{gc}}
{\langle\delta(\sigma_k,+1)\delta(\phi_{k}(\sigma),{\bar E})\rangle_{gc}},
\end{equation}
where ${\bar E}$ denotes one of all possible values of $\phi_{k}(\sigma)$. 
In the microcanonical ensemble,  the energy per site $e=E/V$ 
is held fixed. 
Assuming the equivalence between the 
grand-canonical and microcanonical ensembles, we get the following expression
\begin{equation}
e^{-2\beta {\bar E}}
= \frac{\langle\delta(\sigma_k,-1)
\delta(\phi_{k}(\sigma),{\bar E})  \rangle_{mc}}
{\langle\delta(\sigma_k,+1)\delta(\phi_{k}(\sigma),{\bar E})\rangle_{mc}},
\label{eqmc1}
\end{equation}
which allows us to obtain the temperature $T$  with 
respect to the microcanonical ensemble. A similar procedure can be performed 
for the Potts model. By choosing two particular states $k^{*}$ and
 $k^{**}$ (ranging from $0$ to $q-1$) with respective transition 
$k^{*}\leftrightarrow k^{**}$ and the state function 
$h(\sigma)= \delta(\sigma_k,k^{*}) \delta(\phi_{k}(\sigma),{\bar E})$  
we have,  by appealing to the equivalence of ensembles, that
\begin{equation}
e^{-\beta {\bar E}}
= \frac{\langle\delta(\sigma_k,k^{*})
\delta(\phi_{k}(\sigma),{\bar E})  \rangle_{mc}}
{\langle\delta(\sigma_k,k^{**})\delta(\phi_{k}(\sigma),{\bar E})\rangle_{mc}},
\label{eqmc2}
\end{equation}
where $\phi_{k}(\sigma)$ is given by 
$\phi_{k}(\sigma)=J\sum_{\delta} 
(\delta_{k^{**},\sigma_{k+\delta}}-\delta_{k^{*},\sigma_{k+\delta}})$. 
 Since  the above formulae does not specify the dynamics, they are
valid for different classes of microcanonical algorithms.

\section{Numerical Results}
Numerical simulations have been performed in a square lattice with
$L^{2}$ sites.
The  microcanonical dynamics  is composed of two parts. In the first part, 
a given site of the lattice is  randomly chosen and 
its spin is changed to one of its all possible values. In the second part, 
two sites of the lattice, also randomly chosen, have theirs spins 
interchanged. The above dynamics are  accepted only if the total 
energy remains unchanged. 
It is worth mentioning that the actual MC algorithm
is quite different from those studied
in Refs. \cite{fiore6}, where both energy and magnetization are
strictly conserved. Here the
particle moves are accepted only when the total energy does not change.
The number of species (spins) is not necessarily conserved. 

By applying the logarithm on both sides of Eq. (\ref{eqmc1}), we get 
the following expression (written in units of $J$ and $k_B$)
\begin{equation}
\ln R_{n}=-\frac{2n}{T},
\label{eqmic3}
\end{equation}
where the right-hand of Eq. (\ref{eqmc1}) was written as $R_n$. 
The possible values of the quantity  
${\bar E}=\phi_{k}(\sigma)=J\sum_{\delta}\sigma_{k+\delta}$ 
are given by $nJ$, where $n$  
takes the values $n=-4,-2,0,2,4$, for the Ising model and 
$n=-4,-3,...,3,4$, for 
the BC model. Thus, from the above, by calculating $\ln R_n$ numerically for 
all possible values  of $\phi_{k}(\sigma)$, the temperature  is 
extracted from the inverse of the slope 
of Eq. (\ref{eqmic3}). A similar procedure 
is done for the Potts model, where we have 
\begin{equation}
\ln R_{n}=-\frac{n}{T},
\label{eqmic4}
\end{equation}
where $n$ assumes the values $-4,-3,...,3,4$ for all values of $q$.

\begin{figure}
\centering
\includegraphics[angle=270,scale=0.35]{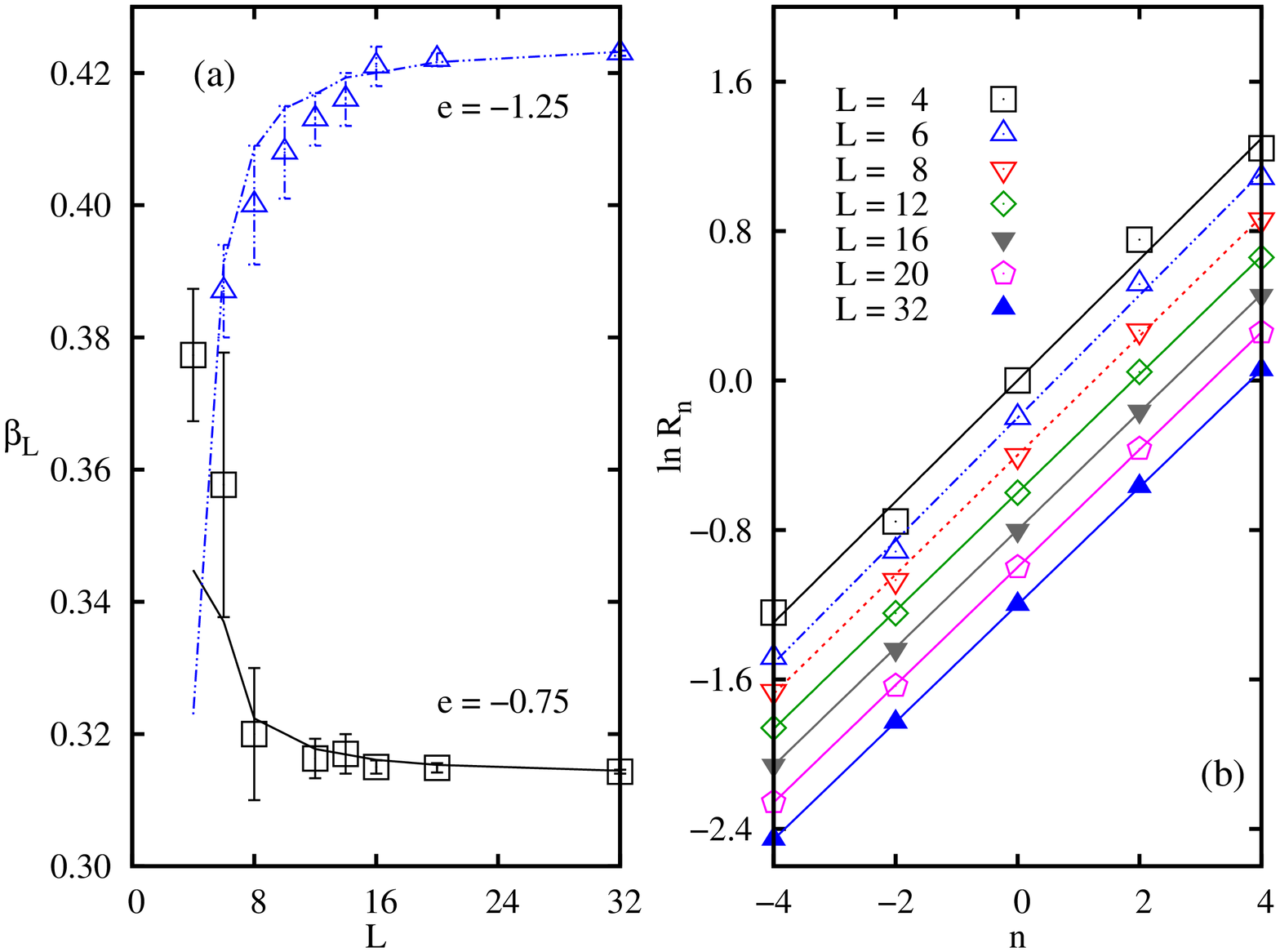}
\caption{Inverse of temperature 
$\beta_L$ versus  $L$ for different $e$'s for the Ising model. 
Symbols and continuous lines 
denote the estimates obtained from the present approach and   the exact 
DOS, respectively.  In $(b)$ we have a mono-log plot of the 
quantity $R_n$ versus $n$ for different $L$ and $e=-0.75$. Curves
have been  shifted in order to avoid overlapping and continuous
lines have been used for better visualization of slopes. }
\label{fig1a}
\end{figure}

In the first analysis, we study the validity of Eqs. (\ref{eqmic3})
and (\ref{eqmic4}) for different $L$, as showed  in  
Fig. \ref{fig1a}$(a)$ and $(b)$ for the Ising model. 
Continuous lines and symbols denote standard and present results, respectively
 and temperatures calculated 
for $L=4$ are exact in both cases. 
Comparison between  intensive quantities show that they are
slightly different for the smallest 
$L$'s ($L=4$ and $L=6$). Inspection of part $(b)$ 
reveals us that in these cases, the quantity $\ln R_n$
is not linear in $n$, and hence  Eq. (\ref{eqmic3})  does not hold. 
The non validity of Eq. (\ref{eqmic3}) for extreme small $L$
is exemplified by evaluating  the numerator and denominator 
of Eq. (\ref{eqmc1}) for $e=-1.25$ and $L=4$.  
Since the number of configurations is small, both quantities are
zero for $n=0$, whereas for $n=4$ ($n=-4$) the numerator (denominator) 
is null.
By increasing $L$   the number of configurations 
becomes large in such a way that the linear dependence between
$\ln R_n$ and $n$ is achieved. Only in this
regime we can evaluate $T$ from Eq. (\ref{eqmic3}).
 In practice,  estimates become equivalent 
for rather small
system sizes. For example, for the Ising model   
 the difference between estimates is in the third decimal level 
for $L=8$.

\begin{figure}
\centering
\includegraphics[angle=270,scale=0.35]{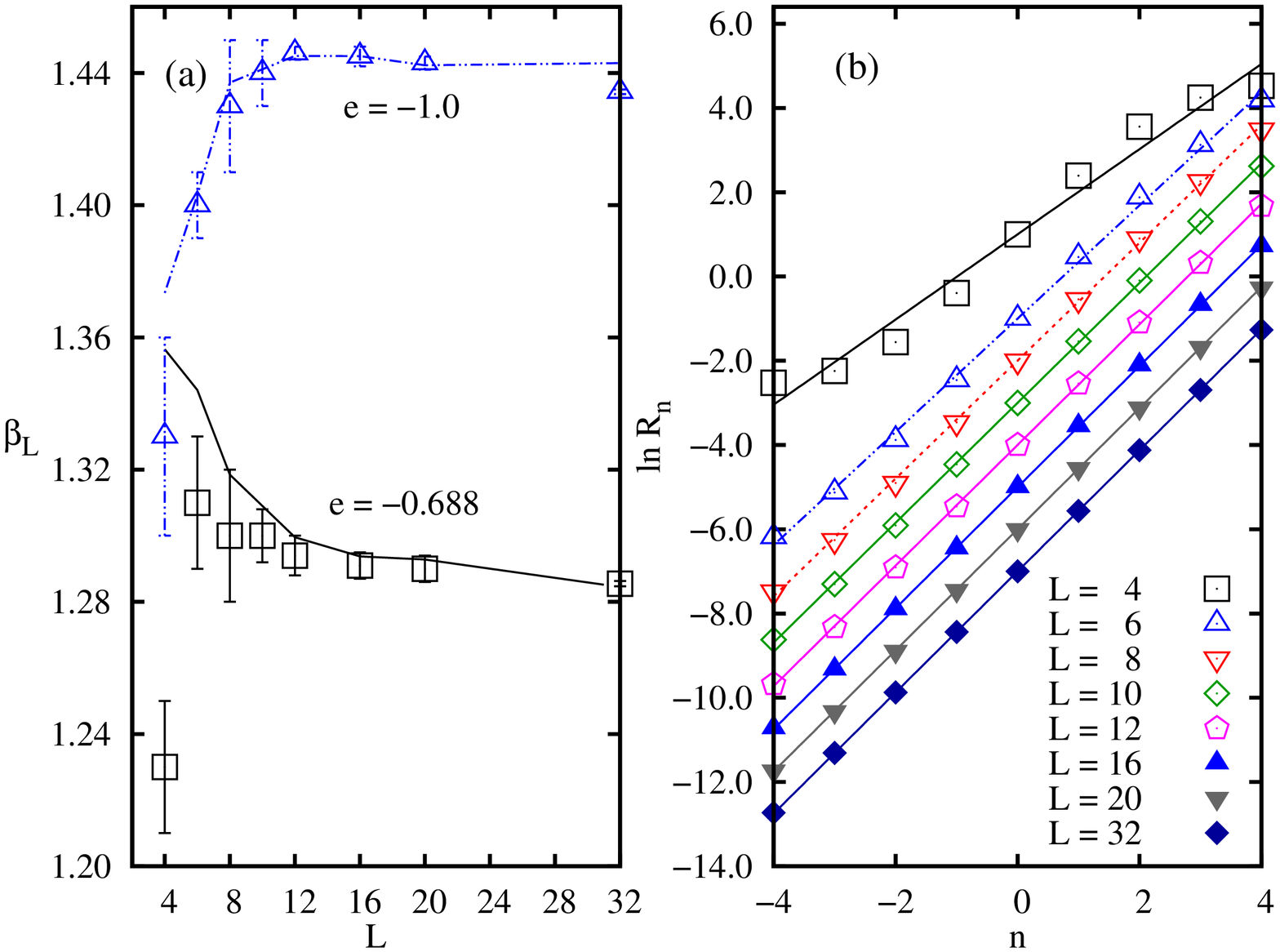}
\caption{Inverse of temperature 
$\beta_L$ versus  $L$ for different $e$'s for the $q=10$ Potts model. 
Symbols and continuous lines 
denote the estimates obtained from the present approach and   the exact 
DOS, respectively.  In $(b)$ we have a mono-log plot of the 
quantity $R_n$ versus $n$ for different $L$ and $e=-1.25$. Curves
have been  shifted in order to avoid overlapping and continuous
lines have been used for better visualization of slopes. }
\label{fig1b}
\end{figure}

Similar conclusions are verified for the other models, as exemplified in
 Fig. \ref{fig1b}$(a)$ and $(b)$ for the $q=10$ Potts model. As in
the Ising model,  Eq. (\ref{eqmic4}) is not hold for
small $L$'s (exemplified in part ($b$) for $e=-1.25$), 
which becomes equivalent to $(\frac{\partial s}{\partial e})_L$ for larger 
(but still rather small) $L$'s. 
It is worth remarking  that due to the small number of configurations
and the discretization of energy, both procedures 
are not precise in the limit of extreme low energies.

\begin{figure}
\centering
\includegraphics[angle=270,scale=0.35]{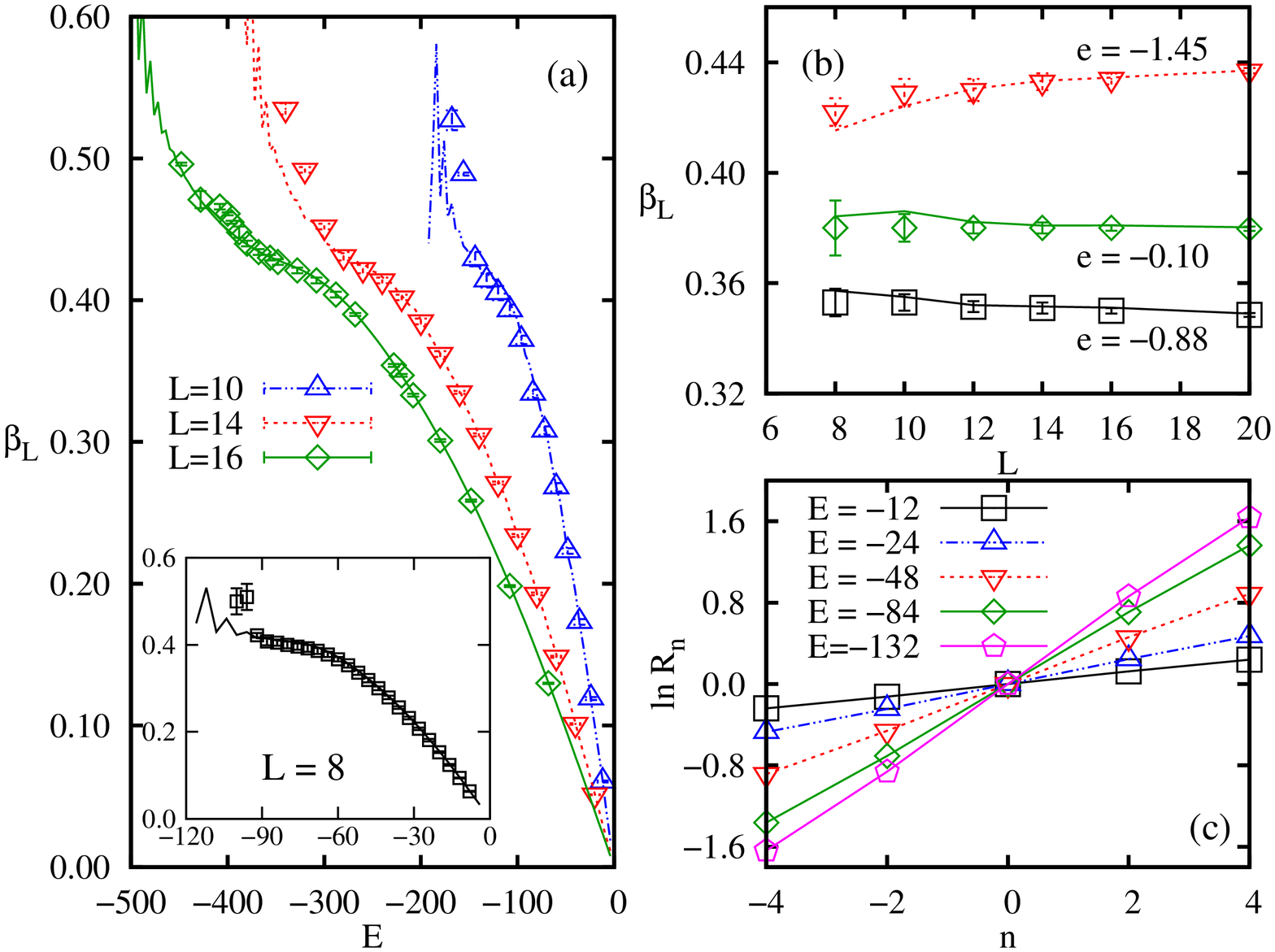}
\caption{Inverse of temperature 
$\beta_L$ versus total energy $E$ for the Ising model and 
different system sizes $L$. Symbols and continuous lines (for $(a)$  and $(b)$) 
denote the estimates obtained from the present approach and   the exact 
DOS, respectively. Inset shows the convergence
of methods already achieved for $L=8$. In $(b)$ we have the $\beta_L$ versus 
$L$ for some 
values of $e=E/V$. In $(c)$ we have a mono-log plot of the 
quantity $R_n$ versus $n$ for $L=10$ and different values of $E$. Slopes
of straight lines give $\beta_L$. In this case, continuous                        
lines have been used for better visualization of slopes. }
\label{fig1}
\end{figure}

Once established the regime of
convergence of methods, we extend  the previous analysis
for the whole range of energy.
In  Fig. \ref{fig1} $(c)$, we plot the   
$\ln R_n$ as function of $n$ for several values of $E$ and $L=10$. 
Note that all curves are linear  
and cross at $(0,0)$, which gives $H=0$ for all energies and temperatures,  
in consistency with results 
 by Beale \cite{beale}, where the DOS 
was enumerated for $H=0$. 
In Fig. \ref{fig1}$(a)$ we plot  $\beta_L$ versus 
the total energy $E$ for different  $L$. In order 
to avoid data overlapping we choose to plot $E$ instead of $e=E/V$. 
The results for the Ising model show an excellent agreement 
between estimates of $\beta_L$  for all system sizes (part $(b)$). 
In addition, we have also compared (not shown) both schemes at 
the ferromagnetic-paramagnetic second-order phase transition.
The pseudo-critical temperature $\beta_{L}^{*}$ may be
estimated 
by the peak in the specific heat $C$ (obtained
from the energy numerical differentiation). 
The deviation between $\beta_{L}^{*}$ and its
asymptotic value  $\beta_c=\frac{1}{2}\ln (1+ \sqrt{2})$ 
(obtained here) agrees very well with exact estimates by Beale \cite{beale}. 

\begin{figure}
\centering
\includegraphics[angle=270,scale=0.35]{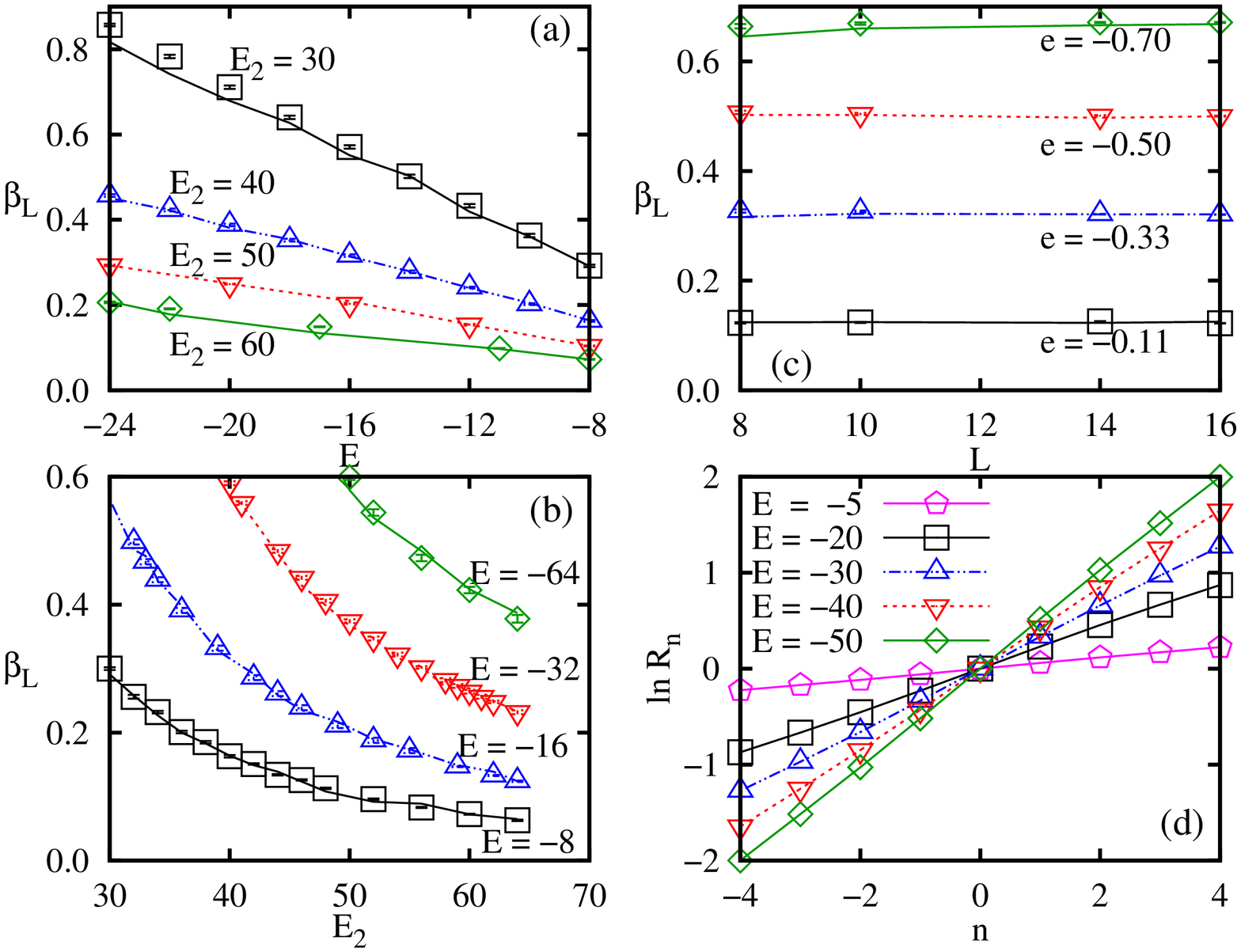}
\caption{In $(a)$ and $(b)$ we plot the 
inverse of temperature $\beta_L$ as a function of $E$ 
($E_2$) for the BC model with $E_2$ ($E$) fixed and $L=8$.
Symbols  and continuous lines                                
denote estimates obtained from the                                                 
present approach and from Wang-Landau  method, respectively.
In $(c)$  we show 
the dependence of $\beta_L$ on $L$ for 
different values of $e$ and $e_2=2/3$. In $(d)$  we  plot   
 $\ln R_n$ versus $n$ for  $E$ for $e_2=2/3$ and $L=10$.
Slopes            
of straight lines give $\beta_L$. In this case, continuous                         
lines have been used for better visualization of slopes.}
\label{fig2}
\end{figure}

%
Further,  we extend the previous analysis to the Blume-Capel model. 
In similarity with the WL model \cite{claudio}, numerical simulations were 
performed for fixed $E=\sum_{(i,j)}\sigma_i\sigma_j$ and 
$E_2=\sum_{i}\sigma_{i}^{2}$.  In practice, $E_2$ fixed
implies that the number of spins $0$ is conserved. 
The one-flip part is restricted to only spins 
$\pm 1$. In Fig. \ref{fig2}$(a)$ we plot $\beta_L$ versus $E$ for 
different $E_2$, whereas in the graph $(b)$, we analyze the dependence
of   $\beta_L$ on $E_2$ for $E$ fixed. 
The dependence  on  $L$  is also showed in Fig. \ref{fig2}$(c)$. 
Note again a very good agreement between both approaches, 
even for small system sizes,  supporting once more
the equivalence between Eqs. (\ref{twl}) and  (\ref{eqmc1}).  

\begin{figure}
\centering
\includegraphics[angle=270,scale=0.35]{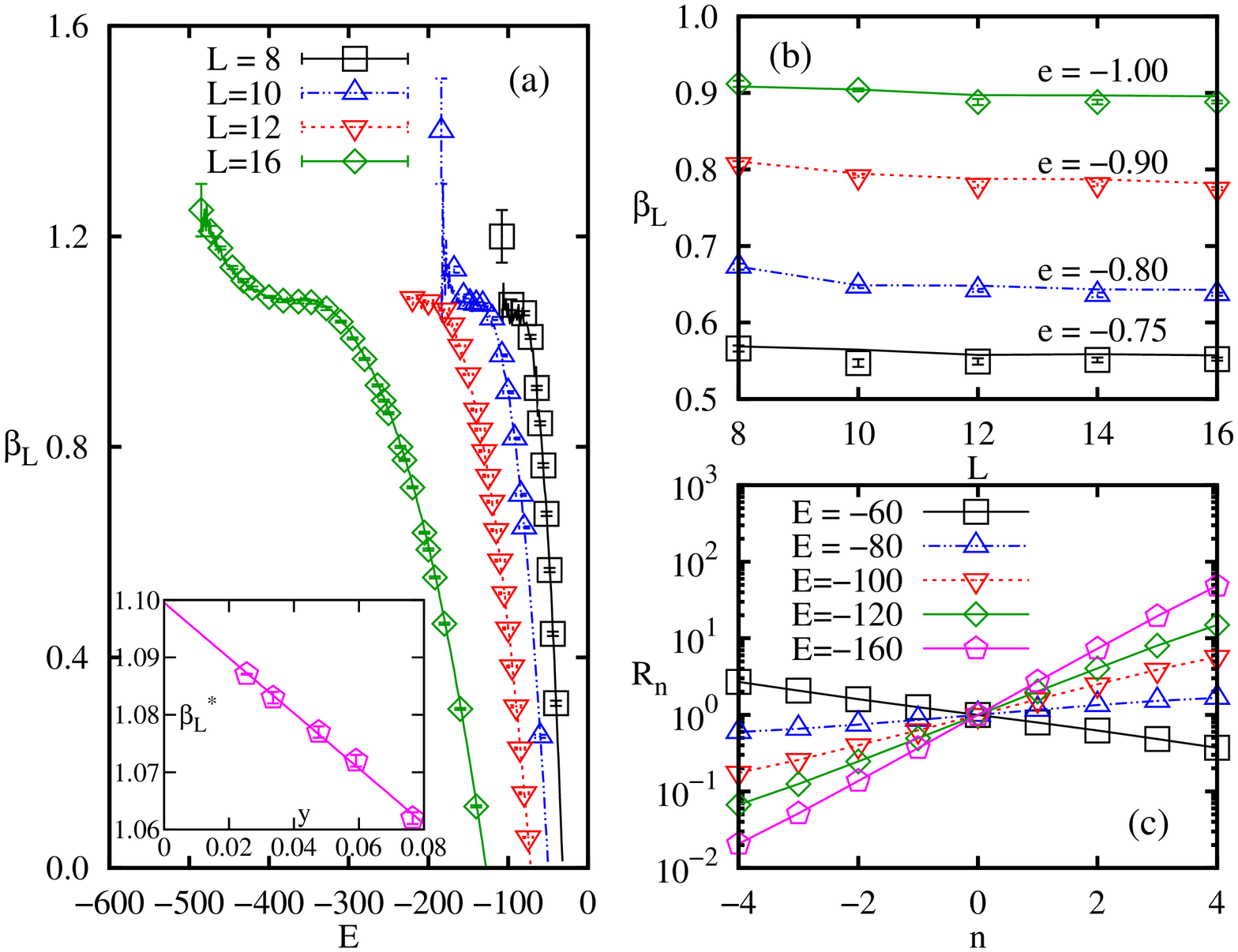}
\caption{In $(a)$, we plot the  inverse of temperature $\beta_L$ 
versus $E$ for several  $L$ in the $q=4$ Potts model. 
Symbols and continuous lines denote estimates obtained 
from present approach and WL, respectively. In
 $(b)$ we plot the dependence of $\beta_L$ on $L$ for 
different values of $e$. In $(c)$ we 
plot the $\ln R_n$ versus $n$ for different energies and $L=10$. 
Slopes  of straight lines give $\beta_L$. In this case, continuous lines have
been considered for better  visualization of slopes.
In the inset  we plot 
 the inverse of the temperature $\beta_{L}^{*}$ in which
specific heat presents a
 maximum  versus $y\equiv \frac{(\ln L)^{3/4}}{L^{3/2}}$. 
The straight line has linear coefficient $\beta_c=1.0982(4)$.}
\label{fig3}
\end{figure}

Now we take the Potts model in kind. As in the Ising and BC models,
it also presents ferromagnetic-paramagnetic phase transitions,
exactly located at $\beta_0=\ln(1+\sqrt{q})$. 
For $q \le 4$, it is second-order 
which becomes first-order for $q>4$. The  case $q=4$  presents
remarkable features, including  
 logarithmic scaling corrections \cite{mayor,fernandez,sokal} 
and  a double peak probability distribution \cite{fukugita}, 
hence an interesting case to be considered.
In Figs. \ref{fig3}$(a)$ and $(b)$ we 
evaluated $\beta_L$ for different energies and system sizes 
(relative small $L$'s but sufficient large to imply the validity
of Eq. (\ref{eqmic4})). 
As in the previous examples, 
we have also found an excellent agreement between estimates obtained 
from microcanonical procedures.
In the inset of Fig. \ref{fig3}$(a)$ we plot  the  pseudo-critical 
temperature $\beta_{L}^{*}$,
obtained from the peak in the specific heat $C$.  
For $q=4$, 
the deviation of  $\beta_{L}^{*}$ from its 
asymptotic value $\beta_c$ decays as $y\equiv \frac{(\ln L)^{3/4}}{L^{3/2}}$ 
\cite{fernandez,sokal}, where we 
found (by using this scaling law) the estimate 
$\beta_c=1.0982(4)$,  in excellent agreement
with the exact value $\beta_c=\ln (1+\sqrt{q})=1.0986123...$.

\begin{figure}
\centering
\includegraphics[scale=0.4]{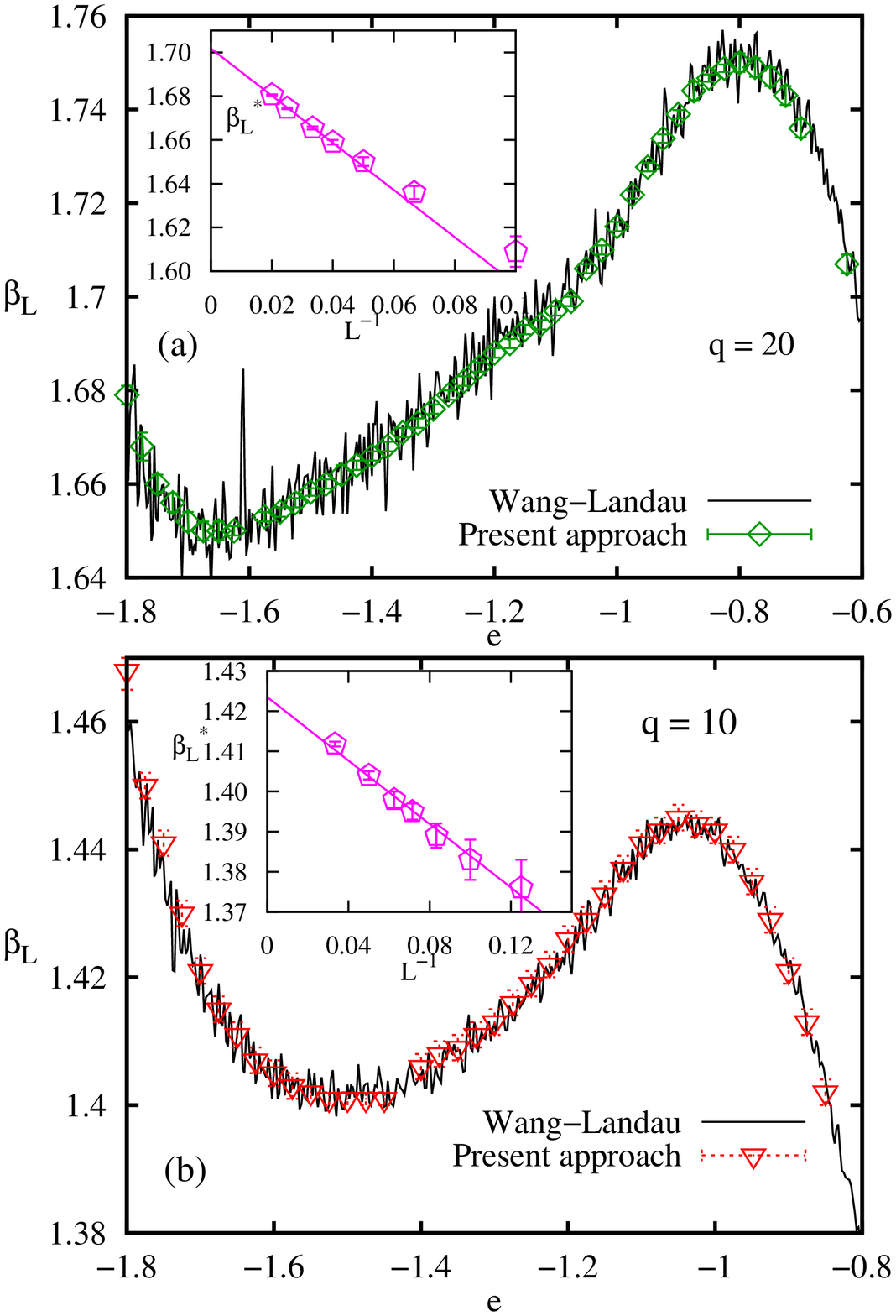}
\caption{Inverse of temperature $\beta_L$ versus $e$ 
for the $q=10$ Potts model   for $L=16$ (top) 
and $L=20$ (bottom). Symbols and continuous 
lines correspond
to the estimates obtained from the present approach and WL, respectively.
In the inset, we plot the  dependence of the minimum   $\beta_L^{*}$ 
versus $L^{-1}$, respectively. 
The straight line has linear coefficient $\beta_0=1.4267(7)$. }
\label{fig4}
\end{figure}

In the last analysis, we evaluated the microcanonical 
temperature for the $q=10$ and $q=20$ states Potts model. 
These  are   also very interesting cases because, in 
contrast to all previous ones, they possesses  discontinuous 
transitions  characterized by a S-like structure, a ``loop'' \cite{komura},
hence ideal examples for illustrating the correctness of the present approach. 
Loops for finite systems in the microcanonical ensemble
are due to interfacial effects, in which  
the surface tension behaves as $1/L$ \cite{gross,komura,janke}.
In contrast, systems simulated in the grand-canonical 
ensemble do not present loops.
In Fig. \ref{fig4} we show the validity of Eq. (\ref{eqmc2})
by plotting $\beta_L$ versus $e$ for $q=10$ (part $(a)$) 
and $q=20$ (part $(b)$) for $L=20$.
As in all previous cases, we also have a good agreement between 
the temperatures, for
both values of $q$. However, in contrast with our results,
estimates obtained from the WL method
presents large fluctuations, even using
the adaptive windows improvement,  taking the flatness criterion
of 90$\%$ for the convergence of $H(E)$ 
and evaluating the mean DOS over 10 different seeds.
On the other hand, our estimates
become more precise as $L$ increases. 
For smaller system sizes  they are less accurate 
(already taking the $L$ for which methods are equivalent),
despite the accordance with estimates from the WL method. This can
be understood in the following: Since 
the number of configurations for fixed $E$ is very large and   
the right side of Eq. (\ref{eqmc2})
is evaluated from only two possible spins,
averages become less precise for small $L$. By increasing $L$,
the number of sites   with above chosen spins are larger and therefore,
the averages becomes more precise. 
This is an interesting point,   since on the contrary 
to the WL, under the present
approach $\beta_L$ 
becomes more precise by increasing $L$. 
In the  inset of each figure, we plot 
the dependence of the minimum $\beta_L^{*}$ on $L^{-1}$, where
the straight line have linear
coefficients $\beta_0=1.4267(7)$ and $\beta_0=1.702(2)$,
which agrees very well with the exact value $\beta_0=\ln (1+\sqrt{10})=
1.42606...$ and $\beta_0=\ln (1+\sqrt{20})=1.699669...$.

\section{Conclusions}
In this paper we have clarified the fundamental issue of a method proposed
at Ref. \cite{fiore6} 
which uses a grand-canonical relationship for calculating microcanonical
quantities. 
The study was carried out from the direct comparison with the standard
definition of microcanonical temperature. 
A detailed analysis for  three different 
lattice models yielding first and second-order phase transitions 
sizes was undertaken.
Our results show that although methods are not
equivalent for extreme small  system sizes, they 
converge for relative small $L$'s
(in practice, our definition 
 becomes equivalent to $(\frac{\partial s}{\partial e})_L$ 
for $L=8$). Not only the estimates for finite
systems were found to be equivalent, but also the thermodynamic
limit transition points. 
The further contribution  exploited its advantages.
Besides the generality and easy implementation,   thermodynamic
quantities are precisely evaluated
from rather short simulations.
Other advantages of the method concerns 
 its low computational cost and not requiring a criterion
for achieving convergence of results.
Once the equivalence has been verified
in distinct situations, we believe that the 
present approach may offer a rather cheap method
 for simulating  more complex
systems, such as    lattice models with continuous variables \cite{domany},
spin-glasses and polymer  systems  \cite{landaubook}.
This will be the   subject of  ongoing work.

\section*{ACKNOWLEDGMENT}
We acknowledge M. G. E. da Luz and Mauricio Girardi for
critical readings of this manuscript and  researcher grant from CNPQ.



\end{document}